\documentclass[showpacs,amsmath,amssymb,epsfig,wrapfig,floatflt,shadow,boxedminipage]{revtex4}
\usepackage{graphicx}
 
\begin{document}

\title{Student understanding of rotational and rolling motion concepts}

\author{Lorenzo G. Rimoldini and Chandralekha Singh}
\email{clsingh@pitt.edu}
\affiliation{Department of Physics and Astronomy, University of Pittsburgh, Pittsburgh, Pennsylvania}

\date{\today}

\begin{abstract}
We investigated the common difficulties that students have with concepts related to
rotational and rolling motion covered in the introductory physics courses.
We compared the performance of calculus- and algebra-based introductory physics 
students with physics juniors who had learned rotational and rolling motion
concepts in an intermediate level mechanics course.
Interviews were conducted with six physics juniors and ten introductory students
using demonstration-based tasks.
We also administered free-response and multiple-choice questions to a large number
of students enrolled in introductory physics courses, and interviewed six additional
introductory students on the test questions (during the test design phase).
All students showed similar difficulties regardless of their background, and
higher mathematical sophistication did not seem to help acquire a deeper
understanding.
We found that some difficulties 
were due to related difficulties with linear motion, while others were tied specifically
to the more intricate nature of rotational and rolling motion.
\end{abstract}       


\maketitle

\section{Introduction}

One of the goals of physics education research is to identify sources of student difficulties in
learning physics and to devise and assess novel curricula and pedagogy that may reduce the 
difficulties~\cite{laws,zollman,alan,fuller,meltzer,lillian}.
Investigation of student difficulties related to various physics concepts
is important for devising instructional strategies to reduce or eliminate the problems.
Previous investigations have documented difficulties in various introductory and
advanced physics courses~\cite{lillian,jim,clement,energy,quantum}, but only few examine
the student understanding of rotational motion and notions associated with it
\cite{ortiz}. This study focuses on the difficulties in learning 
rotational and rolling motion concepts covered in a typical introductory physics course.
In particular, we investigate the understanding of torque, moment of
inertia, rotational energy, and rolling motion by interviewing 
students enrolled in the introductory calculus- and algebra-based courses, 
and in a junior level classical mechanics course using Piagetian style demonstration-based tasks~\cite{piaget}.
We also compare the interview findings with response to the multiple-choice and free-response 
questions on these topics administered to large introductory physics classes.
In general, we found a good agreement between the findings of the interviews and
written tests, and the large population of students involved in the study (as described in Sec.~II) 
suggests that the difficulties related to rotational and rolling motion discussed here are common. 

\section{Method}

Our investigation was composed of two parts.  The first consisted of demonstration-based
interviews with individual students (lasting about 60 minutes each), in which students
were asked a set of questions about lecture-demonstration based tasks.
Once students made their initial predictions, they were asked to perform
the demonstration and reconcile any differences in their
expectations and observation. We employed a ``think-aloud" protocol~\cite{chi}
and asked students to make explicit their thought processes verbally as they reasoned
about the questions. We intervened minimally while they were thinking aloud
but prodded them to ``keep talking" when they appeared to be reasoning silently.
After they had finished answering to their satisfaction, we asked further
questions to clarify points that were not made clear earlier.
Since we were interested in understanding conceptual difficulties, 
qualitative reasoning was considered sufficient.
Think-aloud demonstration-based interviews were conducted with ten students enrolled in
calculus- or algebra-based introductory physics courses (typically science or engineering majors 
who had completed the part of the course dealing with rotational and rolling motion concepts) and six physics juniors 
enrolled in a classical mechanics course who had recently learned rotational and rolling motion using 
the Lagrangian formulation at the University of Pittsburgh.
All students were paid volunteers.
Introductory students participating in the interviews received at least a B grade in their physics midterm 
examination covering rotational and rolling motion concepts.

Although interviews provide an excellent means for probing student reasoning and depth of understanding,
they are time consuming, and only a subset of students can be examined using this method. 
On the other hand, multiple choice questions are easy to grade, compare, and analyze quantitatively, 
but the thought process may not be revealed as clearly by the test answers alone. 
In particular, it is not easy to distinguish correct understanding from right answers for wrong reasons
or ``recall" of concepts and formulas without understanding.
However, well-designed multiple-choice tests administered to a large group of students in conjunction with in-depth 
interviews of a subset of students can be an effective tool for understanding student difficulties. 
Thus, in the second part of our investigation, we designed several free-response questions 
guided by the students' performance in the demonstration-based interviews,
and administered them to students in introductory physics classes. 
After gaining further insight, we devised a 
multiple-choice test, which required concise justifications for each answer. 
The design of the alternative answer choices in the test was based upon the common difficulties 
that were identified in the interviews and free-response questions.
The multiple-choice test was not meant to be comprehensive and cover all topics involved with rolling
and rotational motion, but it focused on important concepts underlying the demonstration-based interview tasks (see Table~\ref{table1}).
We revised the test in several iterations with the help of five faculty members at the University of 
Pittsburgh, while administering it to students in several introductory physics
courses and interviewing six introductory physics students individually on the test questions. 
About 3000 students were involved in the entire study, including the period when test questions
were being revised.
The item analysis was performed for every version of the test to assess how well the distractor choices worked.
The final version of the multiple-choice test consisted of 30 questions, and it was administered to 652 students
from seven calculus- and algebra-based introductory physics courses (including one honor class).  
The reliability index $\alpha$ for the multiple-choice test for different classes ranged from 
$0.68$ to $0.82$ (which is considered good by the standards of test design~\cite{nitko}), and the
point biserial discrimination coefficients were between 0.2 and 0.7 for all questions.
Due to lack of class time, only 11 test questions 
were administered to 17 students in an intermediate level mechanics class (included in Appendix~B). 

\begin{table}
\begin{center}
\begin{tabular}{cc}
\hline
\hline
Concepts &
Multiple-choice questions \\
\hline
Moment of inertia & 3, 4, 20, 24 \\
Rotational kinetic energy & 1, 2, 3, 21 \\
Angular speed/velocity & 5, 13, 22, 25\\
Angular acceleration & 5, 11, 12, 24, 29\\
Torque& 5, 9, 10, 11, 12, 22, 23, 26, 27, 28, 30\\
Rolling (relative motion) & 6, 7, 8, 16\\
Rolling (role of friction and other parameters)& 14, 15, 17, 21\\
Sliding/tumbling cube on an inclined plane & 18, 19\\
\hline
\hline
\end{tabular}
\end{center}
\caption{The different concepts covered and the questions in the multiple-choice test that addressed them (see Appendix B).}
\label{table1}
\end{table}

\section{Demonstration-based Interview Tasks}

In this section, we present the interview questions, and the concepts investigated through them.
In the next sections, we report on student responses to the interview and written questions.

\begin{figure}
\begin{center}
\includegraphics[height=1.7in]{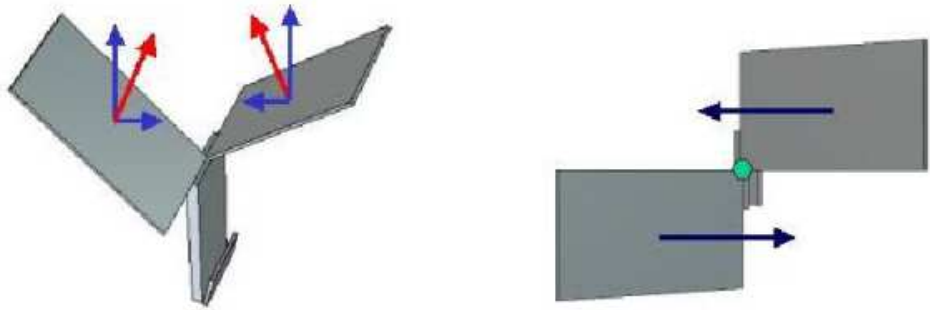}
\caption{A schematic diagram of the paper ``helicopter'' or ``rotor" 
with the drag force and its components along the vertical and horizontal directions. 
The perspective from top (on the right-hand side) only shows the force component that produces the torque to make the paper helicopter rotate.}
\label{i}
\end{center}
\end{figure}

The first interview task was about a paper rotor or ``helicopter"
shown in Fig.~\ref{i} (the device was not referred to as a rotor or helicopter during
the initial prediction phase).  Although the detailed physics of paper helicopter motion could be complicated, students were
only asked questions that could be answered based upon introductory physics concepts.
The interviewer held the paper helicopter vertically with its wings at approximately a $45^{\circ}$ angle with respect to the horizontal, and asked students to 
predict its motion if it were let go from rest. All students correctly predicted that it would rotate as it fell down.
Students were then asked to answer a series of questions assuming that the rotor stayed rigid while falling. The list of questions and expected responses are presented in Appendix~A.
After the initial predictions, students were asked to perform the demonstration and reconcile the differences between
 their predictions and observations.

\begin{figure}
\begin{center}
\includegraphics[height=1.7in]{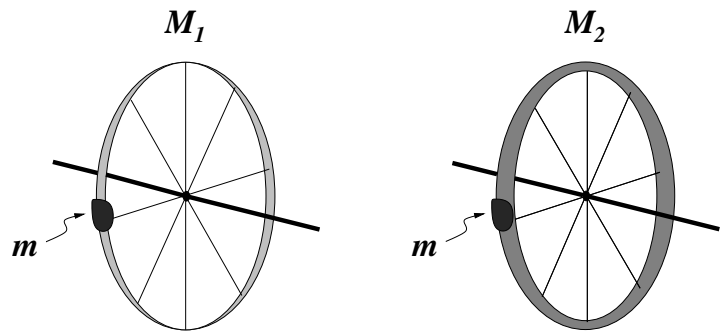}
\caption{The two wheels shown in the figure have the same radius but different masses ($M_1>M_2$)
uniformly spread at the rim of each wheel.
A small piece of putty with mass {\em m} is attached to the rim of each wheel. 
The two systems are shown right after the piece of putty is fastened to the rim.}
\label{ii}
\end{center}
\end{figure}

The second demonstration task posed during the interview involved a comparison of two identically shaped wheels with the same
radius $R$ but with different masses $M_1$ and $M_2$ uniformly spread at the rim.
The wheels were suspended on horizontal axles and could freely rotate about the axles.
Students were asked to sketch and compare qualitatively the
graphs of the angular velocity $\omega$ as a function of time for both wheels when
a small piece of putty of mass $m$ was attached to the rim of each wheel, as shown in 
Fig.~\ref{ii}. They were specifically asked to compare the period, $T$, and the maximum angular speed,
$\omega_{max}$, in the two cases. They were told to assume that the axles were frictionless and they should
ignore air resistance.

We hoped that students would note that when the piece of putty was fastened to the wheel, its 
angular velocity oscillated in time, and that
the wheel with the larger moment of inertia had a longer period but a smaller $\omega_{max}$.
The initial (maximum) torque is $\tau=mgR$ about a point on the central axis due to the piece of putty 
right after it is attached to the rim as shown in Fig.~2. It
causes the wheel to rotate (say) in the clockwise direction. As the piece of putty goes
down, the potential energy of the system consisting of the piece of putty, the wheel, and the Earth decreases, while the 
rotational kinetic energy increases. At the lowest point, the magnitude of the
angular velocity reaches its maximum value $\omega_{max}$. 
As the wheel continues to rotate clockwise, the sign of the torque (and angular acceleration) becomes opposite to $\omega$, so that the piece of putty rises up on the other side slowing down, until the kinetic energy vanishes and the potential energy reaches the maximum value. 
At this point, the magnitudes of the torque and the angular acceleration have their maximum values, and the wheel
starts rotating in the counter-clockwise direction (only now the sign of $\omega$ reverses).
Although the maximum torque $mgR$ is the same for both wheels,
the wheel with the larger mass at the rim has a larger moment of inertia and hence a smaller angular 
acceleration. Therefore the maximum angular velocity of this wheel is smaller (and the period larger).

\begin{figure}
\begin{center}
\includegraphics[height=1.4in]{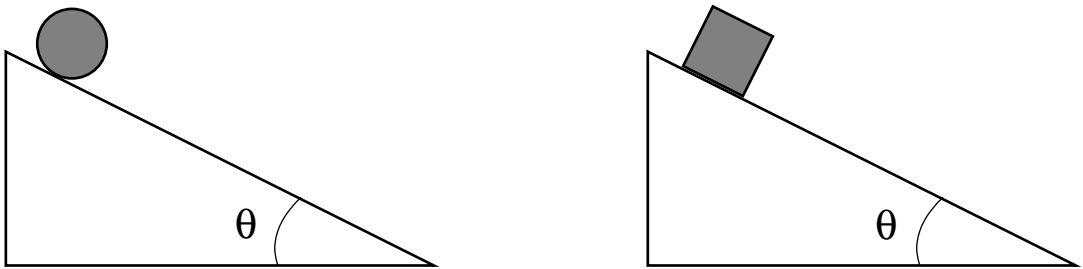}
\caption{An inclined plane with a sphere or a cube released from the top.}
\label{iii}
\end{center}
\end{figure}

The third problem posed during the interviews consisted of predicting the motions of a rigid homogeneous sphere and a cube released from
rest on an inclined plane, and explaining the different factors that affect their motions
(see Fig.~\ref{iii}). After their initial responses, students were specifically asked
to predict the influence of the angle of inclination $\theta$ (if they had not already done so) and
explain the role of friction on the motion, including the limiting cases of infinite and vanishing values
for the coefficient of static friction $\mu_s$. 
We hoped that students would have noticed that in general the sphere would roll without slipping down the inclined plane below
a critical slope (identified by the angle $\theta_c$ with respect to the horizontal), while it would slide for larger angles 
$\theta>\theta_c$. The value of
$\theta_c$ depends on the frictional coefficient $\mu_s$: the higher $\mu_s$ is,
the larger $\theta_c$ becomes.
In the limiting cases of negligible or extremely high $\mu_s$, the sphere would 
slide or roll (without slipping), respectively, for any angle
$\theta\in(0^{\circ},90^{\circ})$. 

This result can be understood, e.g.,  by calculating the torque due to
the frictional force about the center of the sphere. In this case, the line of action of both the
gravitational force and normal force pass through the center of the sphere,
and so they do not produce a torque about the center of the sphere (due to zero lever arms). 
The linear acceleration $a$ of a sphere rolling on an incline is 
$a=g \sin\theta-(F_s/M)$, where $M$ is the mass of the sphere, and 
$F_s$ denotes the static frictional force, leading to rolling when
$\tan\theta<\mu_s$ (which follows from the
condition for not sliding, i.e., $g\sin\theta<g\mu_s\cos\theta$).
The torque equation about the center of the sphere is given by $\tau=-F_sR=I_o\alpha$,
where $R$, $\alpha$, and $I_o$ represent the radius, angular acceleration,
and moment of inertia (about an axis passing through the center) of the sphere, respectively.
For a rolling sphere, the linear and angular accelerations are related through
$a=\alpha R$, and therefore it follows that $F_s=-I_oa/R^2$ (which may be combined with
the expression for $a$ on the incline to solve for $\alpha$).
Another way to think about the motion of the sphere is in terms of a torque about the point
where the sphere touches the inclined plane. For this choice, the lever arms for the frictional force and
normal force are zero, and the gravitational force provides the torque:
$\tau=MgR\sin\theta=I\alpha$, where $I=I_o+MR^2$ (thanks to the parallel-axis theorem).

For a given $\mu$, a rigid homogeneous cube will remain at rest below a critical angle and tumble 
at larger $\theta$. In the limit $\mu\rightarrow 0$, it would slide for all $\theta>0$, while for 
extremely large $\mu$ it would remain at rest for  $\theta<45^{\circ}$ and tumble for $\theta>45^{\circ}$.
The tumbling of the cube can be understood, e.g., in terms of a torque due to the gravitational force
about a point on the edge of the cube touching the inclined surface at the lowest corner. 
For $\theta>45^{\circ}$, the center of gravity is outside of the area of contact, so that the cube tumbles.

The last question posed was a standard textbook problem about a rigid wheel rolling without slipping on a horizontal surface.
Students were asked for the velocity of the top and bottom points at the rim of the wheel with respect to
the ground $\vec v_{pg}$, given that the velocity of the center of mass with respect to the ground was $\vec v_{cg}$. They were
asked to explain their reasoning.
In order to answer this question, students had to understand the concept of relative motion ($\vec v_{pg}=\vec v_{pc}+\vec v_{cg}$). 
The bottom point of the wheel is at rest with respect to the ground as the wheel rolls because the velocity of
that point with respect to the center of mass $\vec v_{pc}$ is equal and opposite to the velocity of the center of mass with
respect to the ground $\vec v_{cg}$.
The velocity of the top point of the wheel with respect to the ground is $2\vec v_{cg}$ because $\vec v_{pc}=\vec v_{cg}$ for that point.

\section{Student Response to the Interview Questions}

A somewhat surprising and discouraging result of our interviews is that no clear difference emerged in the performance of students
in the lower and upper level courses (though no honor student participated in the interviews); it is interesting
to note that a similar result emerged also from studies involving other topics (e.g., Ref.~\cite{wosilait}).
In general, both groups of students had similar difficulties and were unsure about the same 
aspects of rotational and rolling motion concepts covered in the introductory physics courses.
The higher mathematical sophistication employed in the intermediate mechanics course appeared not to help physics juniors 
acquire a deeper understanding of the concepts covered in the introductory courses.  
In all interview questions students were asked to start with a physical setup related to lecture-demonstration tasks. 
It appeared difficult for them to instantiate the ``high-level'' tools they had learned in the upper-level course to those situations.  
The fact that many physics courses do not succeed in teaching students to reason
qualitatively about physical phenomena agrees with other findings~\cite{laws,zollman,alan,fuller,meltzer,lillian}.
Since there was no clear difference in the performance of the introductory and advanced students, 
we do not identify their responses to the interview questions separately in the discussions below.

Another thing to consider before analyzing the results of the interviews
is that the teaching of rotational and rolling motion in an introductory course follows that
of linear motion. Therefore many of the previous difficulties with linear motion compound with difficulties 
involved in learning the more complex concepts related specifically to rotational and rolling motion.  
For example, we found conservation of energy and relative motion to be poorly understood concepts that can 
be related to specific difficulties in understanding rotational and rolling motion.

\subsection{Paper helicopter}

Although all of the interviewed students predicted that the paper helicopter would start to rotate,
only a few (less than 20\%) could explain why the helicopter, which was initially at rest, would start to rotate.
Most students did not use the concept of torque and talked about forces which
``don't balance'' in order to produce a rotation as in the following excerpt:
\begin{quote}
{\em Student 1}: ...forces with different magnitudes will produce a net force causing rotation....
\end{quote}
The dependence of rotational motion on various parameters was equally difficult for them to explain. 
Most students knew that air resistance (drag force) somehow played a role, but 
they often did not understand the direction in which it acted and {\em how} it was affecting the helicopter motion.
Even when the helicopter wings were inclined at an angle $\theta$ with respect to the horizontal, many believed that
the drag force was completely vertical. Thus they were missing the horizontal component needed to produce the torque.
The following responses pertain to the following question by the interviewer:
``Can you tell me how the air-resistance force will act on these wings?''
It was clear from the responses that students had some notion that air resistance
is important but they had difficulty articulating its effect on the motion of the helicopter.
\begin{quote}
{\em Student 2}: Air resistance always acts vertically...
\end{quote}
\begin{quote}
{\em Student 3}: There's an air coming around...air flow...air resistance...
if there's no air resistance then it probably would not rotate...
but if there's air resistance, the air is moving around...and it makes it rotate, 
because air resistance probably gives like...a force on this. \\
{\em Interviewer}:  In which direction does this force act?\\
{\em Student 3}: There's normal force perpendicular to this surface
[referring to the wings] that pushes up, but this other force
like air or force of air...friction...something...that pushes down on
this thing [pointing at the helicopter] [...] 
I think that they're equal but opposite forces, like Newton's third law.
\end{quote}
\begin{quote}
{\em Student 4}: Air resistance is kind of tricky thing. \\
{\em Interviewer}: Why is it a tricky thing? \\
{\em Student 4}: It's a tricky thing because [it is] not really discussed [laughs],
and from what I know it's all approximation... we don't have good equations
for air resistance.
\end{quote}

The following quotes are typical student responses as
they tried to explain why the paper helicopter started to rotate when let go from rest:
\begin{quote}
{\em Interviewer}: [Shaping a piece of paper into a ball, and asking the
student about its motion while it is falling down...] \\
{\em Student 5}: It will not rotate, it'll just fall down. \\
{\em Interviewer}:  Why? \\
{\em Student 5}: It has to do with...(pauses and thinks)...I'm not quite sure why.    
\end{quote}
\begin{quote}
{\em Student 6}: Can you say like...it's in equilibrium when it falls,
and like...in order to maintain equilibrium it needs to rotate.
\end{quote}
\begin{quote}
{\em Student 7}: You need like...an initial force? 
\end{quote}

\subsection{Rotating wheels}

For the second question posed, difficulties were often due to
the inability to apply the principle of conservation of energy.
Several students could describe the rotation of a wheel with a piece of putty attached, but could not 
justify their response properly.
Even those who noted that the angular velocity of the wheel oscillated (when
the piece of putty was attached) did not invoke the conservation of energy
to explain that the piece of putty had to reach the same initial height on the other side (in the ideal frictionless case) before turning back. 
More ``visual'' concepts like speed were easily recalled, but without invoking
appropriate relations with directly related concepts (like kinetic energy). 
The following discussion shows a student who thinks that the piece of putty can reach the same initial height, but does not invoke the conservation of
energy even when asked explicitly about the physical principle involved:
\begin{quote}
{\em Student 8}: On the way down it gains velocity...the velocity is at a 
maximum when it's down here [pointing at the lowest position], 
then it will lose velocity up to zero
[pointing at the same initial height, but on the opposite side of the wheel]. \\
{\em Interviewer}: What guarantees that it will rise to the same height on the other side?
Is there some principle of physics that you can use to figure it out? \\
{\em Student 8}: Just that the velocity you gain on the way down would be canceled
out by the acceleration upward when it goes up.
\end{quote}

A common difficulty in plotting the angular velocity
of the wheels as a function of time was an oscillatory but
entirely positive curve, sometimes with sharp edges.
These kinds of difficulties while describing a motion through a graph are consistent with findings from
other studies~\cite{PT254}.  
When probed explicitly about whether the sign of the velocity depended on the direction of motion,
most students re-drew the $\omega$ vs time curve correctly.
Once students made their prediction, they performed the
demonstration involving the motion of the wheel with a piece of putty attached. After this, one-fourth of the students
tried to draw an analogy with the motion of a pendulum. While it helped them understand qualitatively the oscillatory
motion of the wheel with the piece of putty attached, it confused them
in comparing the behavior of the two wheels.  In fact, they kept thinking
of a simple pendulum (whose period does not depend on the mass of the bob), and
had difficulty realizing that this result would not apply to their problem.

In our investigation, we were also interested in finding out how much of a hint or assistance was required to guide students through the reasoning.
Therefore after the students had explained the observation to the best of their ability while talking aloud,
we gave them successive hints. We did not find any clear difference between the introductory and advanced
students in terms of the amount of assistance required. 
In general, none of the students could compare correctly the plots of the motion of the two wheels 
without hints---if they noted that there had to be a difference in the period in the two cases, they did not correctly predict the 
angular speed in the two cases and vice versa. 
About 20\% of the students claimed that nothing would be different in the two cases.  
However, a few hints were enough to drive about 80\% of them through the correct reasoning. 
\begin{quote}
{\em Student 9}: I think that [the plot for the other wheel] 
would be exactly the same.  \\
{\em Interviewer}:  Can you elaborate? \\
{\em Student 9}: Because, I mean, there's no...no net force...
it seems like the period and angular velocity will be the same for both, I think. \\
{\em Interviewer}:  So, the fact that the wheel is lighter will not change...\\
{\em Student 9}: No, it doesn't matter at all. \\ 
{\em Interviewer}: Why? \\
{\em Student 9}: (thinks for some time) Maybe it will have one change. [...] I think the period
will be the same, but the amplitude may be higher... [changing his mind...] 
the period is different! \\
{\em Interviewer}: Why do you think the period is different now? \\
{\em Student 9}: The period is different because...this one takes...if 
this one goes around faster, it seems like it will get to that point
at faster velocity. [...] hmmm... [murmuring ``pendulum''] 
...this is all related, like a pendulum... \\
{\em Interviewer}:  Why do you think of a pendulum?\\
{\em Student 9}: Oh, it seems like you've a pendulum going like this,
and I think it doesn't matter what mass is on a pendulum. 
\end{quote}
\begin{quote}
{\em Student 10}: I guess the period would be different [...] I'd  say the
angular velocity would be the same.  \\
{\em Interviewer}: Earlier you said this [the lighter wheel] takes a
shorter time to get up to here [the position after half a period]. \\
{\em Student 10}: Right. \\
{\em Interviewer}: So, if it takes a shorter time... \\
{\em Student 10}: Oh..the angular velocity is gonna be much larger for this wheel...
\end{quote}
Many students talked about the difference in ``mass" affecting the angular speed of the wheels.
At the end of the demonstration interview, students
were asked about how the angular velocities and periods
would differ if the masses of the two objects were the same but the mass was distributed differently. Many
did not use the concept of moment of inertia correctly. Some said that they vaguely remember that the
distribution of mass matters but did not remember the exact relation.
Student responses to the multiple-choice questions in the later section also attest to this difficulty.

\subsection{Sphere and cube on an inclined plane}

As for the third interview question, students were unsure about the role of friction in
making a sphere roll and a cube tumble down an inclined plane. 
Several students did not know the meaning of rolling without slipping even though they had just had 
a midterm exam in which concepts related to rolling were covered.
Some believed that the object should roll better if there were no friction.
Others believed that an extremely large coefficient of static friction would imply a Velcro-like effect---i.e., 
no object would be allowed to move at all, even if the inclined plane were almost vertical.
\begin{quote}
{\em Student 11}: If the coefficient of friction is large enough, it
won't move, then. [...]
As long as it  is less than 90 degrees
[referring to the angle between the inclined plane and the
horizontal surface]---at 90 degrees I think
it would fall. [...] 
 Because the coefficient of friction...you don't deal with
that after 90 degrees, but like if it's 89.99 [degrees] and the
coefficient of friction was infinite, it seems like it still sticks there,
because there's still something keeping it attached to it.
\end{quote}
Many students did not differentiate between the concepts of force and torque, and
had trouble with concepts such as the axis of rotation, the lever arm, and
the force involved in producing the torque for the tumbling motion.
Less than one-third of the students predicted the tumbling of the cube for large angles, but even those 
who predicted it had difficulties in describing the cause of such a motion, and
noted that their prediction was based only on intuition.
Others did not conceive the tumbling of the cube, and said that if the slope was
high enough, the cube would start sliding without rotating, no matter how large the friction was. 
After performing the demonstration and observing what actually happened, students had difficulty
in explaining the cause of the kind of motion observed. 
After several successive hints and questioning,
most of the students understood that a cube would tumble down for angles
(between the inclined surface and the horizontal plane) greater than $45^{\circ}$.

\subsection{Rolling wheel}

The last question was the textbook example of the motion of a rigid wheel rolling without slipping on a horizontal surface.
None of the students were able to explain the velocity of the top and bottom points relative to the ground. 
One student recalled from memory the correct values (i.e., $2v$ and $0$, respectively),
but could not explain the reason. 
Interviews showed that the following difficulties precluded students from deriving
the speed of the points on the top and bottom of a rolling wheel
with respect to the ground (despite repeated hints): 
\begin{itemize}
\item[(i)]
They were not comfortable with (linear) relative motion concepts;
\item[(ii)]
They did not understand the meaning of rolling without slipping
(i.e., $v_{\mbox{\scriptsize\em center wheel}}=\omega R$).
\end{itemize}
Some students kept confusing the velocities of the top and bottom points relative to the center of mass of the wheel with
the velocities with respect to the ground:
\begin{quote}
{\em Interviewer}:  What are the speeds of the points here and here
[pointing at the top and bottom of the wheel] [...] for a rolling
wheel, with respect to the ground? \\
{\em Student 12}: With respect to the ground these points are moving
at speed $v$ and $-v$: this one is moving at $-v$ [pointing at one of the
two points], this one is moving at $v$ [pointing at the other point].\\
\end{quote}
When students replied incorrectly, the interviewer tried to help their
reasoning by giving successive hints. Occasionally, even numerous hints and demonstrations were not 
enough to enable them to reach the correct conclusions.  Students were generally
very uncomfortable with relative velocity concepts.
They were so used to thinking in terms of one preferred frame of reference that
a different reference frame represented a change of their acquired ``common-sense.'' 
For example, in the excerpt below, after going through step by step reasoning for why the bottom of the wheel should
have zero velocity relative to the ground, a student was asked about the velocity of the top point with respect to the ground:
\begin{quote}
{\em Student 13}: The top point...this point right here [pointing at the
top of the wheel] is moving at speed $v$ with respect to the ground, and where it touches the
ground is not moving at all.
\end{quote}
After a demonstration and prodding by the interviewer whether they had noticed any blurring of the spokes in the top and bottom parts
in a rolling bicycle wheel, a student who did not differentiate the
speeds of the top and center of the wheel finally realized that those two points
had different speeds with respect to the ground, but could not quantify how different they were:
\begin{quote}
{\em Student 14}: Well, actually that [the top point of the wheel]
moves faster [than the center of the wheel]---it covers a larger angle of displacement, 
definitely [...] \\
{\em Interviewer}:  How much faster is this moving? \\
{\em Student 14}:  That speed is faster than $v$, I'd say...I can't tell exactly.
\end{quote}

\section{Response to multiple-choice and free-response questions}

As described in Sec.~II, we administered free-response and multiple-choice questions related to 
rotational and rolling motion to about 3000 introductory physics students.
The purpose was to evaluate the extent to which the interview findings were confirmed by a general student population.
The final version of the multiple-choice test is included in Appendix~B. 
The frequency of student responses for each of the five choices on each question is presented in 
Table~\ref{table2}, which includes the average response of 652 introductory students from seven different courses (including one honor class) who took the final test version, 
along with the responses of 17 upper-level physics students to a subset of 11 questions from the same test.
The introductory students were given 50 minutes to take the full test while the upper-level physics students
who were administered only 11 questions had 20 minutes to answer the questions.
Both in the introductory and upper-level courses, the test counted as part of the course grade, often in the form of a quiz.
The average test scores in nonhonor (introductory and upper-level) classes ranged from $44\%$ to 
$61\%$, while a class of 93 freshman honor students (required to have a QPA of 3.5 or better) scored $75\%$ on an average.
We note that the results for the correct answers listed in Table~\ref{table2} represent only upper limits, because
many students choosing correct answers actually did not provide proper or satisfactory justifications for their choices.\\
Student responses and explanations of the test questions showed that students had common difficulties, which
were consistent with the findings of the individual interviews. Such difficulties could be classified into two
categories: those sharing a common ancestry with linear motion \cite{francis}, 
and those that are uniquely related to the more
instricate nature of rotational motion.
Table~\ref{table1} broadly categorizes the concepts covered in each test question.

{\em Moment of inertia.} 
Student responses to the questions involving moment of inertia $I$ showed that many students were unsure about this concept.
For example, many did not know that $I$ is a function of the mass distribution
about an axis, and that the rotational kinetic energy depends on $I$ and not just on the total mass. As shown in Table~\ref{table2},
in question (20), many students thought that $I$ depended
on the angular acceleration of the cylinder. Interviews showed that this type of difficulty was partly
due to students' unfamiliarity with $I$. In particular,
students would never claim that, for linear motion, the mass of an object depended upon its acceleration.
Student responses to questions related to rotational kinetic energy showed that students had great
difficulty with the exact dependence of the kinetic energy on the moment of inertia and the angular speed of the object.  
The following were typical explanations from students:
``The larger the mass of a wheel is, the greater the rotational energy is." 
``The lighter wheel has more rotational kinetic energy... since it's moving faster."

{\em Torque, angular acceleration and angular velocity.} 
Students also shared common difficulties on questions related to torque, angular acceleration and
angular velocity similar to the findings of the interviews. 
The definition and meaning of torque were unclear to many students and often replaced by the concept of force. 
Students were often unclear about the concept of lever arm, and when asked explicitly in
the test-based interviews, many considered torque and force as equivalent concepts.
For example, one student claimed the following about the situation in which two forces were applied in opposite directions at the ends of a rod:
``The net torque is zero, because the forces would cancel out... canceled out by opposite forces." This was consistent with several
students' written responses to questions (9)--(12).
Interviews also suggested that several students wanted to calculate the torque due to different forces on a rod about different points (not
all points were on the central axis).
When the interviewer reminded students that they were supposed to calculate the torque about a point on the axis 
passing through the center of the rod (and perpendicular to it), some 
explicitly noted that they were not sure about why it would matter whether the torque was calculated about a point on the central axis. 
Further questioning of some interviewed students showed that some of them recalled that the net torque about any point is zero for a rigid object in equilibrium, and then claimed 
that one can calculate the contribution to the net
torque on a rigid object due to different forces about different points (even if all those points were not on the same axis).
Student responses to the paper helicopter problem during the interview were
similar to those identified in the written responses. However, students
performed better on question (13) than on the second problem
posed in the demonstration interviews. For example,
choice (c) was chosen by a significantly larger number of students than choice (b). 
This result was consistent with the reaction of students during
the interviews: when students were explicitly probed whether the angular velocity should always be positive, they corrected their
original drawings. Student responses to questions (22)--(30) showed that many of them had difficultly understanding 
that the net torque on the disk+clay system or the wheel+clay system about a point on the axle
was produced only by the weight of the clay. 
Student responses to question (30) showed that even when asked explicitly to select an expression for the net torque on the wheel+clay 
system about a point on the axle, many students felt that the mass of the wheel should also contribute to the
net torque similar to their responses to previous questions.
Also, many students did not distinguish or establish a connection between the net torque and angular acceleration, 
partly because they were not comfortable with the concept of moment of inertia. Although fewer students provided the correct response
to question (30) than several other questions between (22)--(29) which were based on the same principle, those who answered (30) correctly
mostly answered questions (22)--(29) correctly as well. Also, in multipart questions, e.g., (22)--(24), 
students who answered the first question about the net torque on the system about a point on the axle 
correctly were significantly more likely to answer the angular acceleration and the maximum angular velocity questions correctly.

Question (5) revealed a common misconception related to linear motion, i.e., the assumption that the net force on an object should be proportional to its velocity. Options (b) and (d) were the most common distractors.
Students who chose answer (b) had a similar misconception about the rotational motion.
The ``constant net force implies constant velocity'' (and vice versa) misconception
became ``constant net torque implies constant angular velocity.''  The following were typical explanations provided by students: 
``Using the fact that a constant force produces a constant velocity and that
torque is the angular version of force, then constant torque would produce constant angular velocity." 
``If there are torques that are constant, the angular velocity must be constant because the torque is moving it."
Students who chose option (d) were confused whether the net torque was proportional to a change in velocity or change in acceleration.

{\em Relative motion.} 
Similar to the interviews, many questions related to rolling motion probed student understanding of relative motion concepts.
Students had great difficulty distinguishing between the speeds of different points on a rigid wheel with respect to the center of the
wheel or ground. 
Most students did not recognize that the bottom point of a rolling wheel was at
rest with respect to the ground. Typical responses from students included:
``The instantaneous velocity with respect to the ground is always tangent to the rolling circle." and
``The speed of all points should be the same with respect to ground because they are all on the same wheel which is rolling."

{\em Rolling motion and friction.} 
Many rolling motion questions also dealt with the condition for rolling and the roles of friction and other parameters
on the rolling motion. A large fraction of students had difficulty with these questions and
believed friction must slow any kind of motion (including rolling).
In the test-based interviews, several students specifically suggested that for a rolling wheel kinetic friction is relevant. They claimed that 
a constant force must be applied to keep a wheel rolling, otherwise friction 
would slow it down. 
When students who were interviewed were specifically reminded
that the wheel was rigid and they should ignore air resistance, their response was unchanged.
Finally, responses to questions (18) and (19) about the motion of a cube on an inclined plane 
(with different $\mu_s$) revealed results consistent 
with the interview findings described in the previous section. 

Question (14) was the most difficult question on the test for both introductory and upper-level students (apart from the honor class). It showed that students had difficulties in
determining the role of friction in rolling without slipping of a rigid wheel on a rigid horizontal surface in the 
absence of air resistance. Responses of general introductory students to questions (14) and (22) also showed that, unlike honor students, they had  
great difficulty in dealing with idealized situations, e.g., considering objects as rigid, axles as frictionless, and ignoring air resistance~\cite{intuition}.

\normalsize
\begin{table}
\centering
\begin{tabular}[t]{cccccccccccccccc}
\hline
\hline
Answers & & (a) & & & (b) & & & (c) & & & (d) & & & (e) &  \\[0.5 ex]
\hline
Questions & GI & HC & UL & GI & HC & UL & GI & HC & UL & GI & HC & UL & GI & HC & UL \\[0.5 ex]
\hline
 1& 16 & 10 &   & 23 & 16 &   & {\bf 57} & {\bf 73} & {\bf  } & 4 & 1 &   & 1 & 0 &  
\\[0.5 ex]
 2& 39 & 5 &   & {\bf 44} & {\bf 92}& {\bf  } & 7 & 1 &   & 3 & 0 &   & 7 & 2 &  
\\[0.5 ex]
 3& 18 & 27 & 29 & {\bf 45} & {\bf 71} & {\bf 41} & 35 & 2 & 18 & 1 & 0 & 6 & 1 & 0 & 6
\\[0.5 ex]
 4& 13 & 7 & 6 & 22 & 15 & 6 & {\bf 61} & {\bf 76} & {\bf 82} & 3 & 0 & 6 & 0 & 2 & 0
\\[0.5 ex]
 5& 6 & 0 & 0 & 24 & 14 & 6 & 15 & 4 & 18 & 26 & 27 & 24 & {\bf 28} & {\bf 55} & {\bf 53}
\\[0.5 ex]
 6& 2 & 0 & 0 & {\bf 37} & {\bf 45} & {\bf 35} & 8 & 5 & 0 & 48 & 49 & 65 & 5 & 1 & 0
\\[0.5 ex]
 7& 7 & 3 &   & {\bf 49} & {\bf 86} & {\bf  } & 30 & 8 &   & 9 & 1 &   & 5 & 2 &  
\\[0.5 ex]
 8& 34 & 6 & 35 & 6 & 3 & 0 & 3 & 1 & 6 & 5 & 2 & 0 & {\bf 52} & {\bf 88} & {\bf 59}
\\[0.5 ex]
 9& 7 & 1 &   & 7 & 3 &   & {\bf 63} & {\bf 80} & {\bf  } & 10 & 11 &   & 14 & 5 &   
\\[0.5 ex]
10& 15 & 1 & 6 & 19 & 19 & 29 & 8 & 5 & 0 & {\bf 52} & {\bf 74} & {\bf 65} & 6 & 1 & 0
\\[0.5 ex]
11& 5 & 0 &   & 1 & 2 &   & 3 & 0 &   & {\bf 82} & {\bf 95} & {\bf  } & 9 & 3 &  
\\[0.5 ex]
12& 2 & 2 &   & 8 & 3 &   & {\bf 78} & {\bf 87} & {\bf  } & 5 & 2 &   & 7 & 6 &  
\\[0.5 ex]
13& 5 & 3 &   & 18 & 12 &   & {\bf 67} & {\bf 85} & {\bf  } & 7 & 0 &   & 2 & 0 &  
\\[0.5 ex]
14& 2 & 2 & 6 & {\bf 19} & {\bf 61} & {\bf 18} & 46 & 14 & 18 & 5 & 14 & 6 & 28 & 9 & 53
\\[0.5 ex]
15& 16 & 6 &   & 11 & 4 &   & 11 & 3 &   & 2 & 0 &   & {\bf 62} & {\bf 87} & {\bf  }
\\[0.5 ex]
16& 21 & 14 &   & 4 & 0 &   & {\bf 60} & {\bf 82} & {\bf  } & 4 & 1 &   & 12 & 3 &  
\\[0.5 ex]
17& 16 & 12 & 6 & 3 & 1 & 0 & {\bf 52} & {\bf 81} & {\bf 53} & 26 & 5 & 41 & 2 & 1 & 0
\\[0.5 ex]
18& {\bf 69} & {\bf 89} & {\bf 70} & 25 & 7 & 12 & 1 & 0 & 18 & 2 & 0 & 0 & 3 & 4 & 0
\\[0.5 ex]
19& 31 & 41 & 18 & 13 & 9 & 12 & 14 & 14 & 35 & {\bf 38} & {\bf 34} & {\bf 24} & 4 & 2 & 12
\\[0.5 ex]
20& 2 & 2 &   & 3 & 2 &   & {\bf 71} & {\bf 85} & {\bf  } & 18 & 9 &   & 6 & 2 &  
\\[0.5 ex]
21& 15 & 1 &   & 9 & 1 &   & 3 & 1 &   & {\bf 66} & {\bf 95} & {\bf  } & 7 & 2 &  
\\[0.5 ex]
22& 6 & 2 &   & {\bf 46} & {\bf 73} & {\bf  } & 28 & 20 &   & 19 & 3 &   & 2 & 2 &  
\\[0.5 ex]
23& 13 & 14 &   & 25 & 22 &   & 3 & 1 &   & 4 & 0 &   & {\bf 56} & {\bf 63} & {\bf  } 
\\[0.5 ex]
24& {\bf 33} & {\bf 61} & {\bf  } & 32 & 24 &   & 3 & 0 &   & 4 & 0 &   & 28 & 15 &  
\\[0.5 ex]
25& {\bf 28} & {\bf 58} & {\bf  } & 32 & 21 &   & 4 & 2 &   & 34 & 18 &   & 2 & 1 &  
\\[0.5 ex]
26& {\bf 79} & {\bf 96} & {\bf  } & 11 & 2 &   & 3 & 0 &   & 2 & 0 &   & 6 & 2 &  
\\[0.5 ex]
27& {\bf 50} & {\bf 78} & {\bf  } & 14 & 4 &   & 7 & 6 &   & 2 & 0 &   & 27 & 12 &  
\\[0.5 ex]
28& 47 & 28 &   & 14 & 8 &   & 3 & 0 &   & 1 & 0 &   & {\bf 35} & {\bf 64} & {\bf  }
\\[0.5 ex]
29& 19 & 9 &   & {\bf 60} & {\bf 75} & {\bf  } & 3 & 0 &   & 1 & 1 &   & 17 & 15 &  
\\[0.5 ex]
30& {\bf 35} & {\bf 67} & {\bf 70} & 37 & 12 & 24 & 12 & 9 & 0 & 12 & 2 & 0 & 4 & 10 & 6
\\[0.5 ex]
\hline 
\hline
\end{tabular}
\caption{Multiple-choice questions (included in Appendix~B) were administered to a total of
669 students. The performance of 559 general (calculus- and algebra-based) introductory
nonhonor students (GI) is distinguished from a honor class (HC) of 93 introductory students,
and an upper-level (UL) class of 17 physics majors enrolled in an intermediate mechanics course (who were administered a subset of 11 questions). The table presents the average percentage (rounded 
to the nearest integer) of students selecting the answer choices (a)--(e) for each question of the 
test (bold numbers refer to the correct responses).} 
\label{table2}
\end{table}

\section{Summary}

Our investigation involving individual student interviews using demonstration-based tasks and written tests 
suggests that students have many common difficulties about rotational and rolling motion concepts 
covered in the introductory physics courses.
Regardless of their level, students share difficulties with
fundamental concepts such as torque, moment of inertia, rotational energy, rolling, and the related role of friction. 
Some of the problems identified here propagate from similar ones with linear motion, while others are
due to the more intricate nature of rotational and rolling motion.
Many students acted as though they were applying rotational principles to real physical situations for the first time.
They seemed uncomfortable in predicting the outcome of experiments
and in reconciling the differences between their initial predictions and observations by taking advantage of the tools they had learned.
They in general had difficulty applying the principle of conservation of energy and relative velocity concepts.
Instructional strategies that focus on improving student understanding of rotational and rolling motion concepts should take into account these difficulties.
The multiple-choice test that we have developed can help assess the effectiveness of strategies to improve student understanding of
these concepts.

\section*{ACKNOWLEDGMENTS}

We wish to thank Jim Stango for his care in setting up the demonstration apparatus, and all faculty members 
(Y.\ Choi, R.\ P.\ Devaty, W.\ Goldburg, A.\ Janis, R.\ Johnsen, J.\ Levy, J.\ Mueller, G.\ A.\ Stewart, J.\ A.\ Thompson, M. Vincent, and X.\ L.\ Wu) 
for helping with the design and administration of the free-response and multiple-choice questions.
This work was supported in part by the National Science Foundation.

\vspace{1cm}

\appendix
\section{Follow-up questions about the rigid rotor and the qualitative responses considered adequate}

\normalsize
\begin{itemize}
\item[(1)] Can you explain why it will start to rotate when let go? 
\item[(2)] Can you predict whether this particular rotor (shown to the student) will rotate clockwise or counter-clockwise? Why?
\item[(3)] What will be the effect of increasing the wing angle with respect to the horizontal keeping all the other parameters fixed? 
In particular, what will happen if the wings were completely vertical or horizontal and why? 
\item[(4)] What will be the effect of increasing the length of the wing for a fixed angle keeping all the other parameters fixed (i.e., assume the
total mass and moment of inertia of the rotor are not affected significantly)?
 Why?
\item[(5)] What will be the effect of increasing the relative displacement of the two wings from the central vertical axis
laterally, keeping all the other parameters fixed (students were shown with gesture what was meant by increasing the lateral displacement of the
two wings to make the question clear to them)? Why? In particular, how will the motion change if the wings were not displaced 
with respect to the center?
\item[(6)] How will the magnitude of the linear downward acceleration of the rotor
compare to the magnitude of acceleration due to gravity, {\em g}? Why? 
\end{itemize}

We hoped that students would have noticed that the air resistance or drag force, $F$, acted perpendicularly to the wings,
and since the wings were displaced, the horizontal component
of $F$ caused a torque about the central vertical axis (but no net horizontal force) on the rotor, as shown in Fig.~\ref{i}.
This torque makes the helicopter rotate when let go. The direction in which the helicopter rotates can be predicted by noting
the direction of the net torque (inferred from the way in which the wings are displaced).

The dependence of the torque on the wing angle (with respect to the horizontal)
is nonmonotonic.  In the extreme case when the wings are completely
vertical, the drag force perpendicular to the wing surfaces is negligible.
On the other hand, for perfectly horizontal wings the air resistance is maximized, but
no horizontal component of the drag force is generated.
Hence in both cases the torque vanishes, while it reaches a maximum value for 
some intermediate angle (which is not trivial to compute, considering that 
the terminal speed of the helicopter depends on the wing angle too).

Keeping all the parameters fixed, if the wings are made longer but the angle with the horizontal is kept fixed, 
both the vertical and horizontal components of the drag force 
increase due to an increase in the surface area.
If changes in the rotor mass and moment of inertia are negligible 
(i.e., most of the mass is distributed in the central part of the rotor, 
and not in its wings), then a stronger torque leads to a larger terminal 
angular velocity (we note that air resistance suppresses the angular 
acceleration rather quickly, at least for a paper rotor, and then the rotor falls
uniformly while rotating at constant $\omega$). 

The increase in the relative displacement of the two wings from the vertical central axis
keeping all the other parameters fixed increases the torque due to an increase in the lever arm. In particular,
if the wings are not displaced with respect to the central axis, the horizontal components of the drag forces on the two 
wings are not displaced from each other (the lever arm is zero about the central vertical axis) so that there is no 
torque to start the rotational motion.

Finally, the vertical downward acceleration of the rotor is smaller than $g$ because of the air resistance, 
the turbulences created by the rotational motion, and the rotation of the rotor (in fact,
the gravitational potential energy of the rotor-Earth system as the rotor falls is not converted wholly
to linear kinetic energy but also to rotational kinetic energy).
Once the terminal speed is reached, the vertical acceleration of the rotor becomes zero.
If students mentioned any one of the above qualitative reasons for why the vertical downward acceleration 
of the rotor was smaller than $g$, it was considered sufficient.

\newpage
\section{Multiple-Choice Test}
\noindent
\underline{Instructions:} FOR ALL 30 QUESTIONS:\\
$\bullet$ Fill in only \underline{one} answer for each question on the Scantron sheet.\\
$\bullet$ Provide an explanation for your choice under each question.\\
$\bullet$ All objects are homogeneous (made of the same material throughout) and perfectly rigid. \\
$\bullet$ Always ignore the retarding effects of air resistance.\\

\normalsize
\begin{enumerate}

\item
Two copper disks (labeled ``A'' and ``B'')
have the same radius but disk B is thicker with four times the mass of disk A.  They spin on frictionless axles.
If disk A is rotating twice as fast as disk B, which disk has more rotational kinetic energy?
\begin{center}
\includegraphics[height=1.5in]{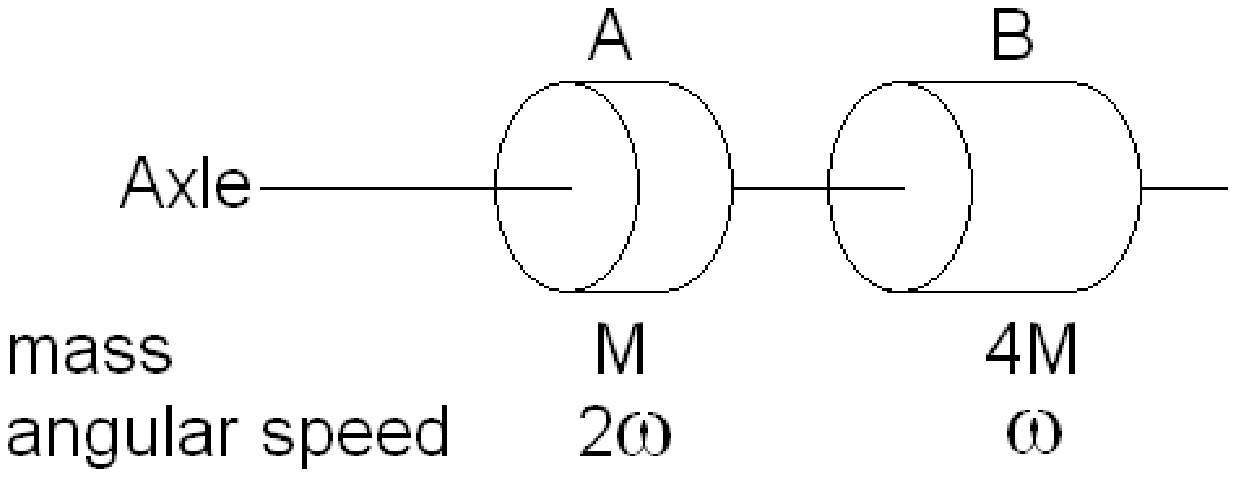}
\end{center}
\begin{itemize}
\item[(a)]
The faster rotating disk A.
\item[(b)]
The thicker disk B.
\item[(c)]
Both disks have the same rotational kinetic energy.
\item[(d)]
It depends on the actual numerical values of the angular speeds of the disks.
\item[(e)]
None of the above.
\end{itemize}
\noindent
{\bf  Explain:}
\vspace{0.1in}

\item
On a horizontal floor, a rigid sphere \underline{rolls} without slipping 
and a rigid cube \underline{slides} without rotating. Both objects have the \underline{same mass $M$}. 
At a certain instant, the points at the center of mass of the sphere and the cube have the same speed $v$
relative to the floor.
Which one of the following statements is {\em necessarily true} at that instant?
\begin{center}
\includegraphics[height=1.8in]{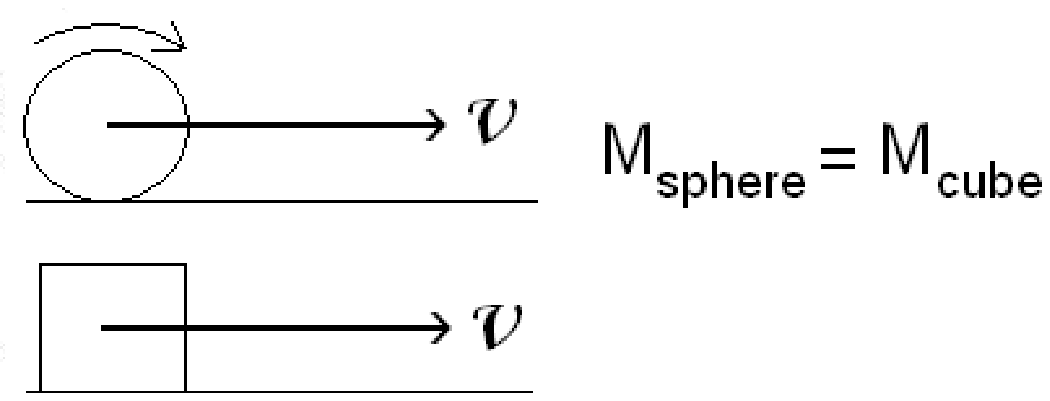}
\end{center}
\begin{itemize}
\item[(a)]
The cube and the sphere have the same total kinetic energy.
\item[(b)]
The cube has a smaller total kinetic energy than the sphere.
\item[(c)]
The work required to stop the cube is greater than that required to stop the sphere.
\item[(d)]
Which object has a larger total kinetic energy depends on the actual numerical value of the mass $M$.
\item[(e)]
None of the above.
\end{itemize}
\noindent
{\bf  Explain:}

\newpage

\item
An aluminum disk and an iron wheel (with spokes of negligible mass) 
have the \underline{same} mass $M$ and radius $R$. They are spinning 
around their frictionless axles with the \underline{same} angular speed as shown.
Which of them has more rotational kinetic energy?
\begin{center}
\includegraphics[height=1.4in]{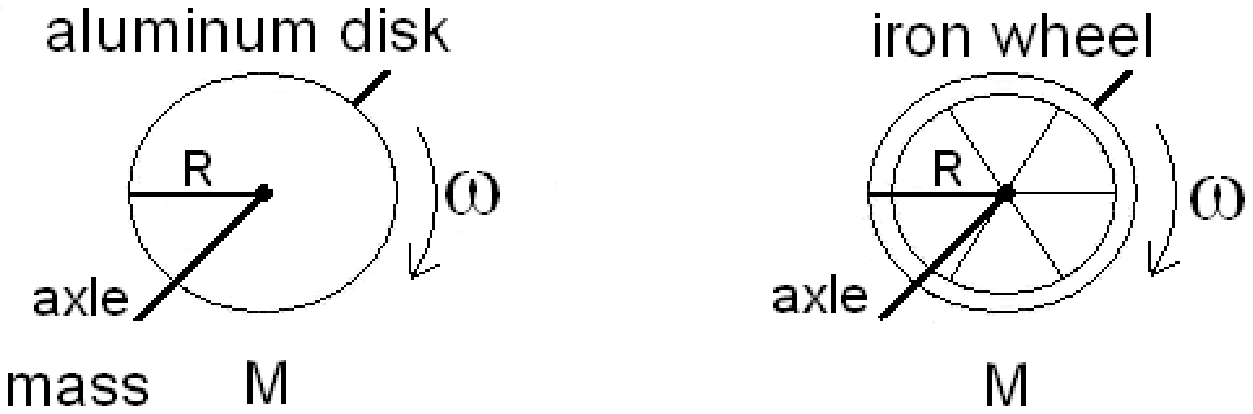}
\end{center}
\begin{itemize}
\item[(a)]
The aluminum disk.
\item[(b)]
The iron wheel.
\item[(c)]
Both have the same rotational kinetic energy.
\item[(d)]
It depends on the actual numerical value of the mass $M$.
\item[(e)]
None of the above.
\end{itemize}
\noindent
{\bf  Explain:}
\vspace{0.1in}

\item
Consider the moment of inertia, $I$, of the rigid homogeneous disk of mass $M$ shown below, about an axis through its center (different shadings only differentiate the two parts of the disk, each with equal mass $M/2$).
Which one of the following statements concerning $I$ is correct?
\begin{center}
\includegraphics[height=1.27in]{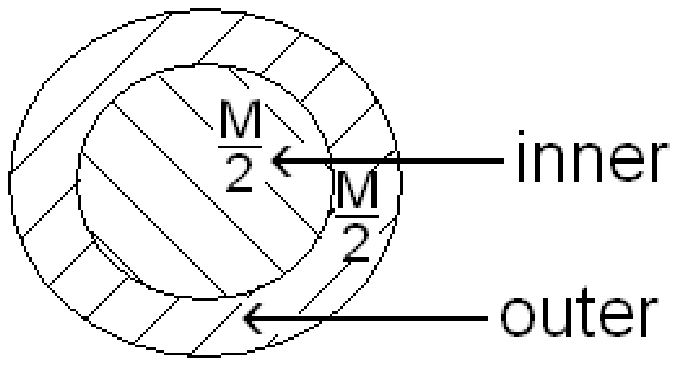}
\end{center}
\begin{itemize}
\item[(a)]
The inner and outer parts of the disk, each with mass $M/2$ (see figure), contribute equal amounts to $I$.
\item[(b)]
The inner part of the disk contributes more to $I$ than the outer part.
\item[(c)]
The inner part of the disk contributes less to $I$ than the outer part.
\item[(d)]
The inner part of the disk may contribute more or less to $I$ than the outer part depending on the actual numerical value of the mass $M$
of the disk.
\item[(e)]
None of the above.
\end{itemize}
\noindent
{\bf  Explain:}
\vspace{0.1in}

\item
Whenever a constant non-zero net torque acts on a rigid object, it produces a
\begin{itemize}
\item[(a)]
rotational equilibrium.
\item[(b)]
constant angular velocity.
\item[(c)]
constant angular momentum.
\item[(d)]
change in angular acceleration.
\item[(e)]
change in angular velocity.
\end{itemize}
\noindent
{\bf  Explain:}
\vspace{0.1in}

\newpage

{\bf {\Large $\bullet$} \underline{Setup for the next three questions}}\\
A rigid wheel of radius $R$ rolls without slipping on a horizontal road.  
The linear velocity of the center of the wheel with respect to the road is $\vec v$ and the angular speed is $\omega$.

\begin{center}
\includegraphics[height=1.8in]{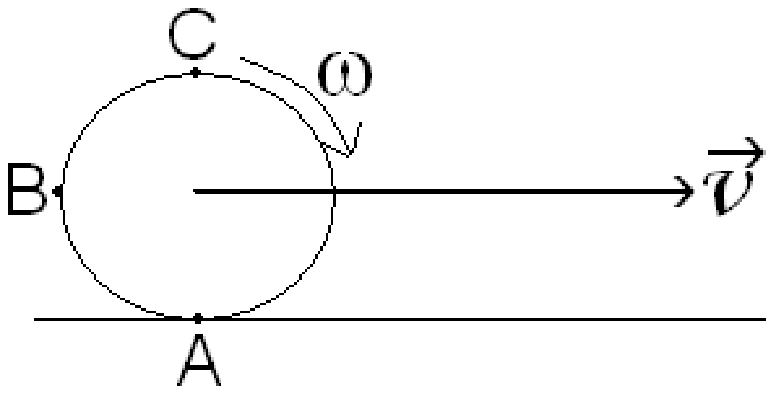}
\end{center}

\item
The direction of the instantaneous velocity of point B (see the figure above) \underline{with respect to the road} is roughly:

\begin{itemize}
\item[(a)] 
$\searrow$
\item[(b)] 
$\nearrow$
\item[(c)] 
$\rightarrow$
\item[(d)] 
$\uparrow$
\item[(e)] 
There is no direction since the instantaneous speed of point B is zero with respect to the road.
\end{itemize}
\noindent
{\bf  Explain:}
\vspace{0.1in}

\item
Which one of the following statements is true about the speed of the points at the rim of the wheel \underline{with respect to the 
{\em road}} (see the figure above)?
\begin{itemize}
\item[(a)]
It is larger for points at the bottom of the wheel than for points at the top of it.
\item[(b)]
It is smaller for points at the bottom of the wheel than for points at the top of it.
\item[(c)]
It is the same for all points, with $v=\omega\, R$.
\item[(d)]
It is the same for all points, with $v=2 \, \omega\, R$.
\item[(e)]
None of the above.
\end{itemize}
\noindent
{\bf  Explain:}
\vspace{0.1in}

\item
Rank order the speeds of points A, B, C at the rim of the wheel
\underline{with respect to the {\em road}}, largest first (see the figure above).

\begin{itemize}
\item[(a)]
$v_A=v_B=v_C$.
\item[(b)]
$v_A>v_B>v_C$.
\item[(c)]
$v_B>v_C>v_A$.
\item[(d)]
$v_C>v_A>v_B$.
\item[(e)]
$v_C>v_B>v_A$.
\end{itemize}
\noindent
{\bf  Explain:}

\pagebreak

{\bf {\Large $\bullet$} \underline{Setup for the next four questions}}\\
The figures in the next four questions show three cases in which a rigid rod of length $2L$ is acted upon by some forces. 
All forces labeled $F$ have the same magnitude.

\item
Which cases have a non-zero \underline{net torque} acting on the rod about its center?
\begin{center}
\includegraphics[height=1.05in]{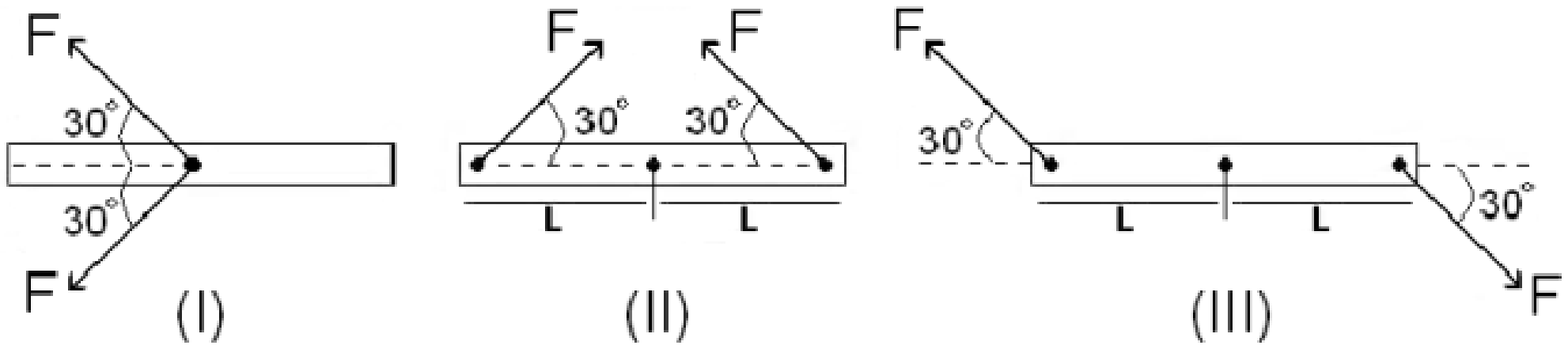}
\end{center}
\begin{itemize}
\item[(a)]
(I) only.
\item[(b)]
(II) only.
\item[(c)]
(III) only.
\item[(d)]
(I) and (II) only.
\item[(e)]
The net torque is zero in all cases.
\end{itemize}
\noindent
{\bf  Explain:}
\vspace{0.1in}

\item
Which cases have a non-zero \underline{net torque} acting on the rod about its center?

\begin{center}
\includegraphics[height=1.05in]{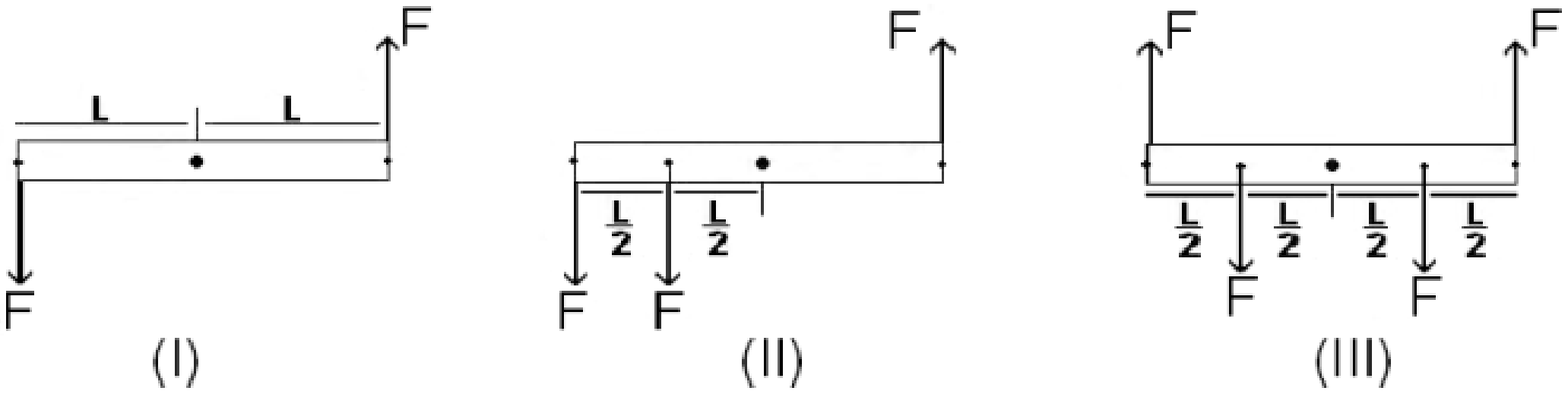}
\end{center}
\begin{itemize}
\item[(a)]
(I) only.
\item[(b)]
(II) only.
\item[(c)]
(III) only.
\item[(d)]
(I) and (II) only.
\item[(e)]
The net torque is zero in all cases.
\end{itemize}
\noindent
{\bf  Explain:}
\vspace{0.1in}

\item
Rank order the three cases according to the angular acceleration of the rod about an axis passing through its center and perpendicular to the
paper.
\begin{center}
\includegraphics[height=1.1in]{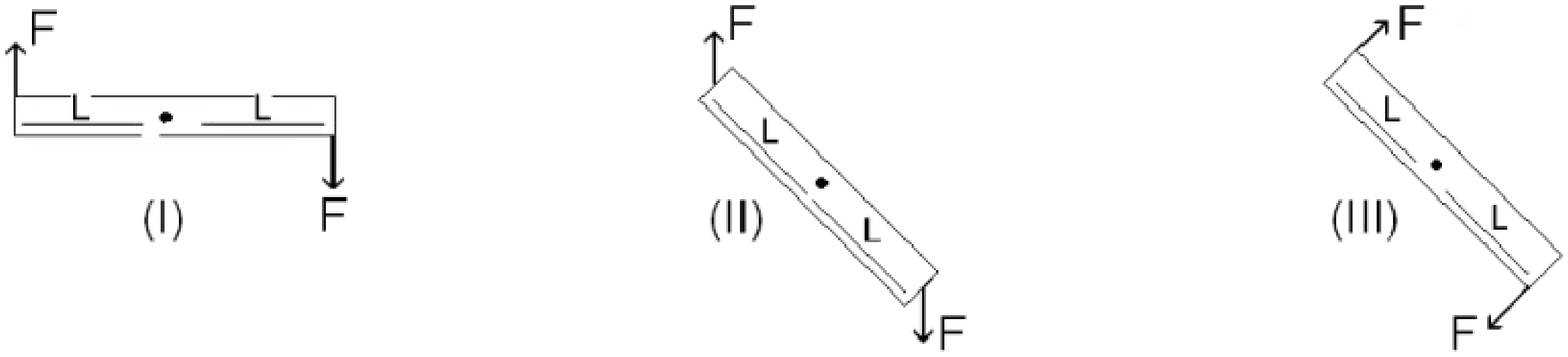}
\end{center}
\begin{itemize}
\item[(a)]
$(III)>(II)>(I)$.
\item[(b)]
$(III)=(II)>(I)$.
\item[(c)]
$(II)>(III)>(I)$.
\item[(d)]
$(I)=(III)>(II)$.
\item[(e)]
$(I)=(II)=(III)$.
\end{itemize}
\noindent
{\bf  Explain:}

\newpage

\item
Rank order the three cases according to the angular acceleration of the rod about an axis passing through its center and perpendicular to the
paper, largest first.
\begin{center}
\includegraphics[height=1.49in]{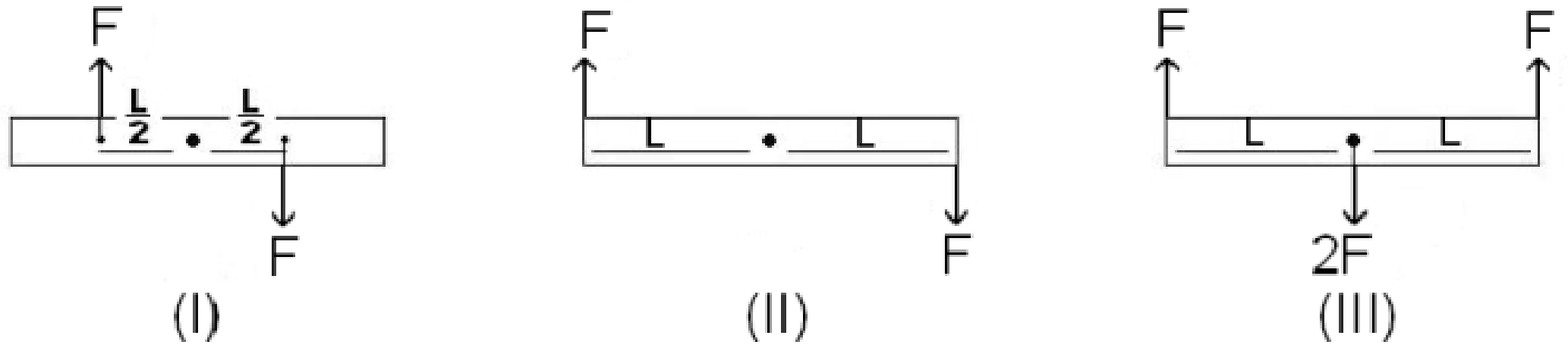}
\end{center}
\begin{itemize}
\item[(a)]
$(I)=(II)>(III)$.
\item[(b)]
$(I)>(II)>(III)$.
\item[(c)]
$(II)>(I)>(III)$.
\item[(d)]
$(II)>(III)>(I)$.
\item[(e)]
The angular acceleration is zero in all three cases.
\end{itemize}
\noindent
{\bf  Explain:}
\vspace{0.1in}

\item
Consider the small-amplitude oscillatory motion of an ideal frictionless pendulum, consisting of
a small mass $M$ attached to a rod of negligible mass.
Ignoring air-resistance, the angular velocity of the mass about the frictionless pivot will be:
\begin{center}
\includegraphics[height=1.1in]{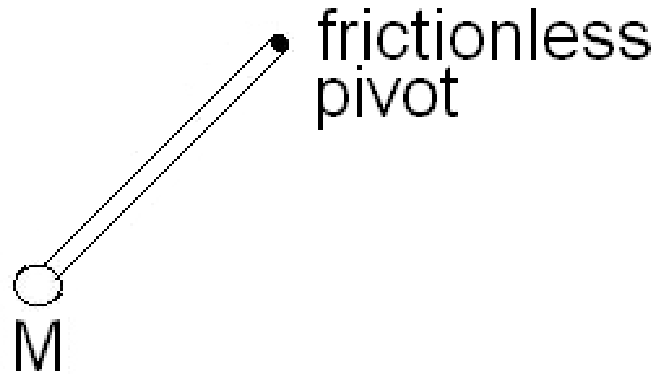}
\end{center}
\begin{itemize}
\item[(a)]
constant in time.
\begin{center}
\includegraphics[height=.5in]{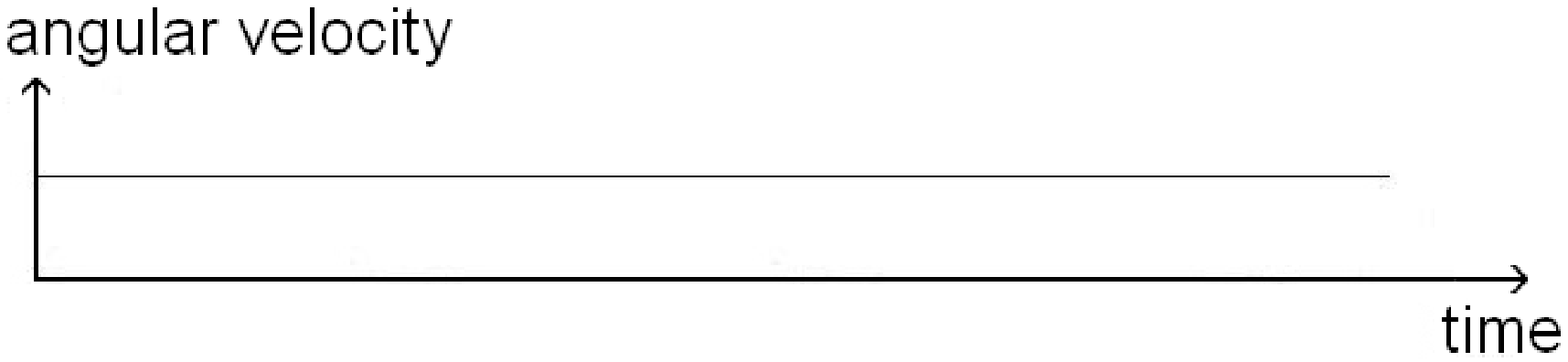}
\end{center}
\item[(b)]
oscillating with fixed amplitude and always having the same sign.
\begin{center}
\includegraphics[height=.5in]{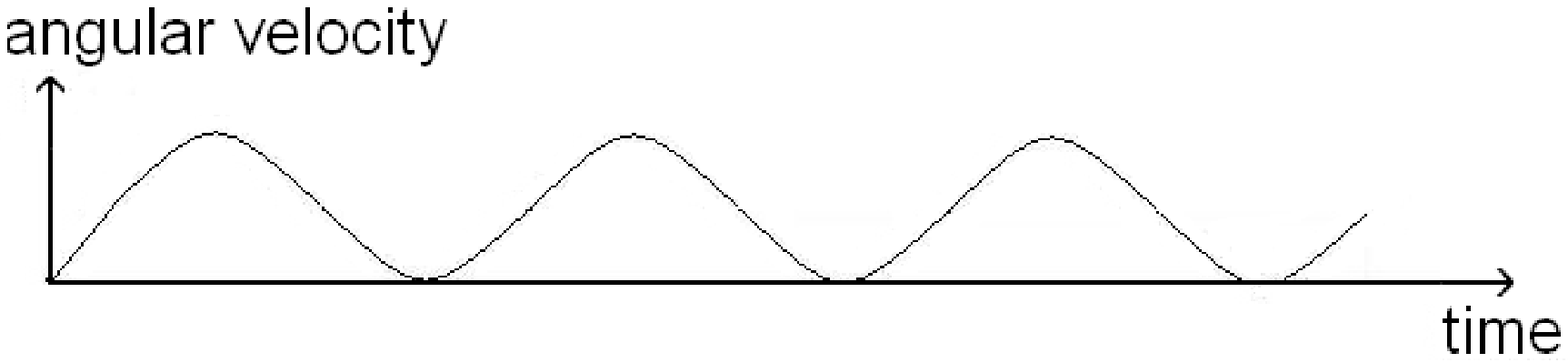}
\end{center}
\item[(c)]
oscillating with fixed amplitude and alternating between positive and negative.
\begin{center}
\includegraphics[height=.5in]{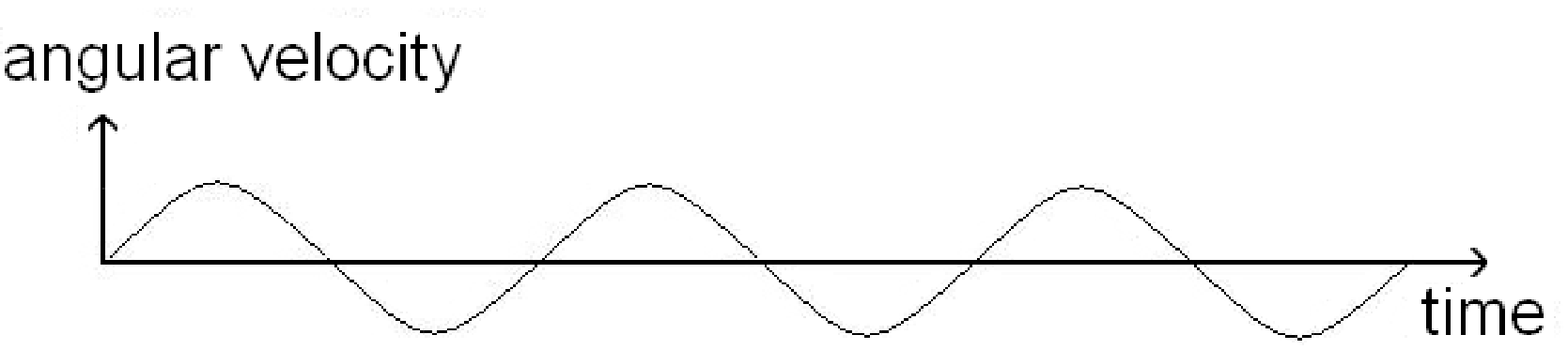}
\end{center}
\item[(d)]
oscillating with decaying amplitude and always having the same sign.
\begin{center}
\includegraphics[height=.7in]{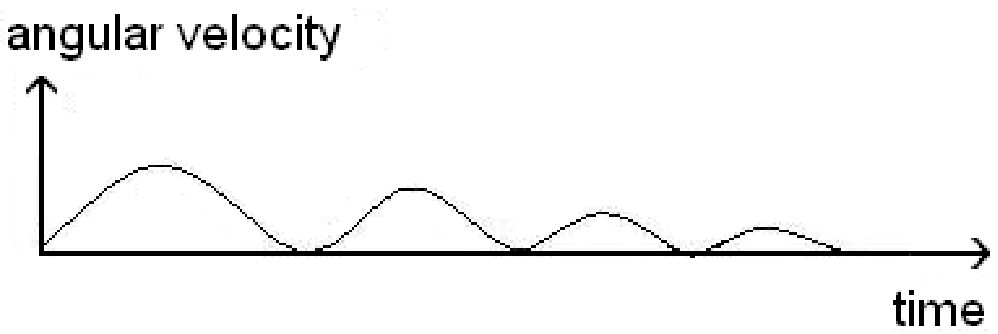}
\end{center}
\item[(e)]
None of the above.
\end{itemize}
\noindent
{\bf  Explain:}

\newpage

\item
Two \underline{identical} rigid marbles roll without slipping across rigid horizontal floors. One rolls on 
a stone floor with coefficient of static friction $\mu_s=0.80$, and the other rolls on a glass
floor with $\mu_s=0.40$. Which marble is slowed down more by friction, and why? Ignore air-resistance.
\begin{itemize}
\item[(a)]
Both marbles are slowed equally because the marbles themselves are identical.
\item[(b)]
Neither marble is slowed by friction because both roll without slipping.
\item[(c)]
The marble rolling on stone is slowed more, because the greater $\mu_s$ makes the force of friction on it greater.
\item[(d)]
The marble rolling on glass is slowed more, because the slippery nature of the glass impedes rolling.
\item[(e)]
It is impossible to answer without knowing the coefficient of kinetic friction $\mu_k$ because the
marbles are moving.
\end{itemize}
\noindent
{\bf  Explain:}
\vspace{0.1in}

\item
If a rigid spherical ball is rolling without slipping down a stationary inclined plane, then 
\begin{center}
\includegraphics[height=1.4in]{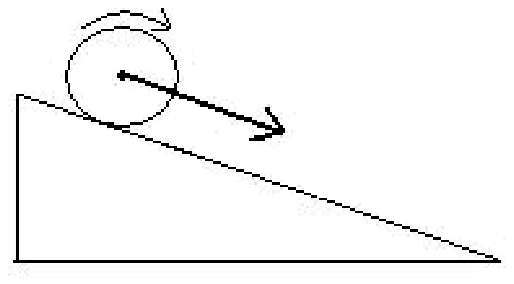}
\end{center}
\begin{itemize}
\item[(a)]
the coefficient of friction between the ball and the plane is precisely equal to 1.
\item[(b)]
the ball has a negligible angular acceleration.
\item[(c)]
the ball has a negligible moment of inertia.
\item[(d)]
the plane slopes steeply, at more than $45^0$ to the horizontal.
\item[(e)]
the static friction between the ball and the plane is not negligible.
\end{itemize}
\noindent
{\bf  Explain:}
\vspace{0.1in}

\item
A bicycle wheel rolls without slipping on a \underline{horizontal} floor.
Which one of the following is true about the motion of points on the rim of the wheel, \underline{relative to the
axis at the wheel's center}?
\begin{center}
\includegraphics[height=1.3in]{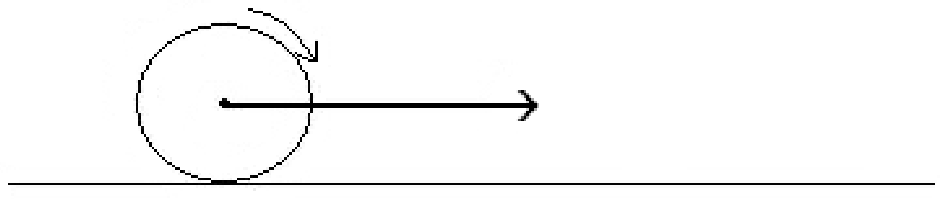}
\end{center}
\begin{itemize}
\item[(a)]
Points near the top move faster than points near the bottom.
\item[(b)]
Points near the bottom move faster than points near the top.
\item[(c)]
All points on the rim move with the same speed.
\item[(d)]
All points have velocity vectors that are pointing in the radial direction toward the center of the wheel.
\item[(e)]
All points have acceleration vectors that are tangent to the wheel.
\end{itemize}
\noindent
{\bf  Explain:}

\newpage

{\bf {\Large $\bullet$} \underline{Setup for the next three questions}}\\
{\bf 
You are to conduct a series of trials.
For each trial, the inclination of the plane is set to an angle $\theta$, ranging from $0^{\circ}$ to $90^{\circ}$, 
and an object is released \underline{from rest} at the top of the stationary inclined plane.
The coefficient of static friction between the object and the inclined plane is $\mu_s$. 
In each case below, predict the observed outcome for the trial.
}\\
In the following questions, `tumble' means `tip over and rotate' and `sliding' means NO tumbling.\\

\item
Case 1: The object is a sphere and $\mu_s=0$:
\begin{center}
\includegraphics[height=.85in]{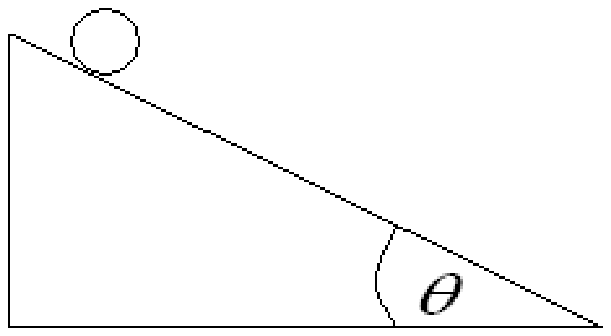}
\end{center}
\begin{itemize}
\item[(a)]
The sphere will roll without slipping for small $\theta$ and slide down only for $\theta$ greater than a certain non-zero value.
\item[(b)]
The sphere will remain at rest for small $\theta$ and roll without slipping only for $\theta$ greater than a certain non-zero value.
\item[(c)]
The sphere will slide down for all $\theta>0^{\circ}$.
\item[(d)]
The sphere will roll without slipping for all $\theta>0^{\circ}$.
\item[(e)]
None of the above.
\end{itemize}
\noindent
{\bf  Explain:}
\vspace{0.1in}

\item
Case 2: The object is a cube and $\mu_s=0$. In this experiment, the angle $\theta$ is varied only between
$0^{\circ}$ and $45^{\circ}$:
\begin{center}
\includegraphics[height=.85in]{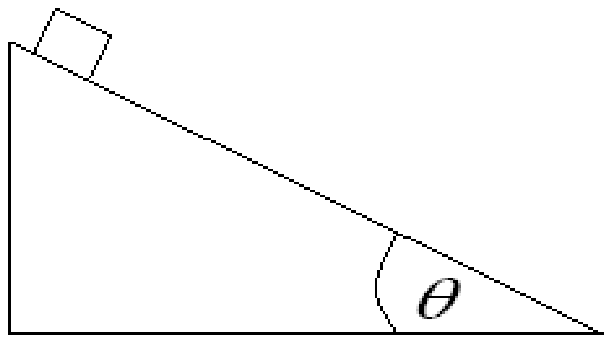}
\end{center}
\begin{itemize}
\item[(a)]
The cube will slide down for all $45^{\circ}>\theta>0^{\circ}$.
\item[(b)]
The cube will remain at rest for small $\theta$ and slide down only for $\theta$ greater than a certain non-zero value.
\item[(c)]
The cube will roll without slipping for all $45^{\circ}>\theta>0^{\circ}$.
\item[(d)]
The cube will tumble down for all $45^{\circ}>\theta>0^{\circ}$.
\item[(e)]
None of the above.
\end{itemize}
\noindent
{\bf  Explain:}
\vspace{0.1in}

\item
Case 3: The object is a cube and $\mu_s$  is very large ($\mu_s>>1$):
\begin{itemize}
\item[(a)]
The cube will remain at rest for small $\theta$ and slide down only for $\theta$ greater than a certain non-zero value.
\item[(b)]
The cube will stay on the incline as if it were glued to it, for all $\theta$.
\item[(c)]
The cube will move relative to the inclined plane only for $\theta=90^{\circ}$.
\item[(d)]
The cube will tumble down only for $\theta>45^{\circ}$.
\item[(e)]
None of the above.
\end{itemize}
\noindent
{\bf  Explain:}

\pagebreak

\item
The moment of inertia of a rigid cylinder 
\begin{itemize}
\item[(a)]
does not depend on the radius of the cylinder.
\item[(b)]
does not depend on the mass of the cylinder.
\item[(c)]
depends on the choice of rotation axis.
\item[(d)]
depends on the angular acceleration of the cylinder.
\item[(e)]
can be expressed in units of kg.
\end{itemize}
\noindent
{\bf  Explain:}
\vspace{0.1in}

\item
You apply a horizontal force, $\vec F$, to the center of a rigid wheel which is \underline{initially at rest} on a 
\underline{horizontal} floor.
If the wheel starts rolling without slipping, then:
\begin{center}
\includegraphics[height=1.in]{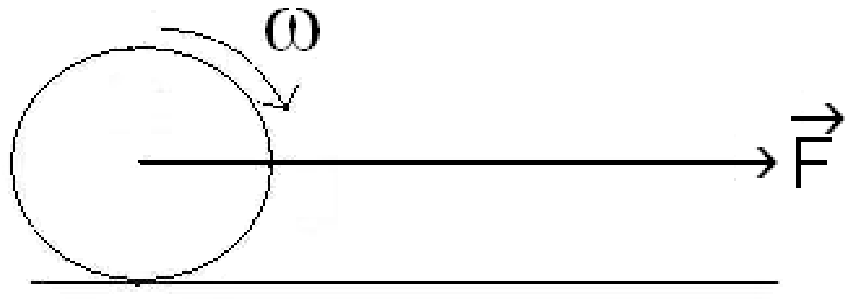}
\end{center}
\begin{itemize}
\item[(a)]
All of its kinetic energy is in rotational form.
\item[(b)]
The coefficient of static friction between the wheel and the floor must be precisely equal to 1.
\item[(c)]
It will roll {\it better} on a frictionless floor.
\item[(d)]
The coefficient of static friction between the wheel and the floor {\it cannot} be zero.
\item[(e)]
None of the above.
\end{itemize}
\noindent
{\bf  Explain:}
\vspace{0.2in}

{\bf {\Large $\bullet$} \underline{Note: In questions 22-30 below, gravitational force should be taken into account.}}\\

\item
A bicycle wheel is free to rotate about a horizontal frictionless axle.
A small lump of clay is attached to the rim as shown in the figure, and the wheel is released from rest.
Which one of the following statements about the motion of the wheel is \underline {true}?

\begin{center}
\includegraphics[height=1.5in]{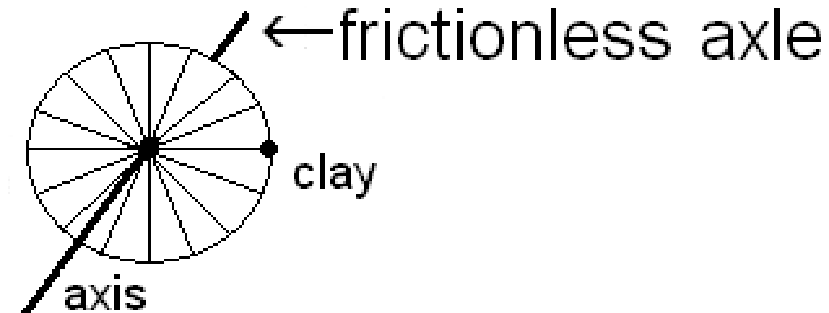}
\end{center}

\begin{itemize}
\item[(a)]
The wheel will remain at rest if the mass of the lump of clay is smaller than a
certain critical value.
\item[(b)]
The wheel will oscillate.
\item[(c)]
The wheel will rotate and come to rest when the lump of clay is at the lowest
possible point.
\item[(d)]
The number of full revolutions (angle $2\pi$) the wheel makes before coming
to rest depends on the mass of the lump of clay.
\item[(e)]
None of the above.
\end{itemize}
\noindent
{\bf  Explain:}

\pagebreak

{\bf {\Large $\bullet$} \underline{Setup for the next three questions}}\\
{\bf 
An aluminum disk and an iron wheel (with spokes of negligible mass)
have the same radius $R$ and mass $M$ as shown below. 
Each is free to rotate about its own fixed horizontal frictionless axle.
Both objects are initially at rest.
\underline{Identical} small lumps of clay are attached to their rims as shown in the figure.}
\begin{center}
\includegraphics[height=1.5in]{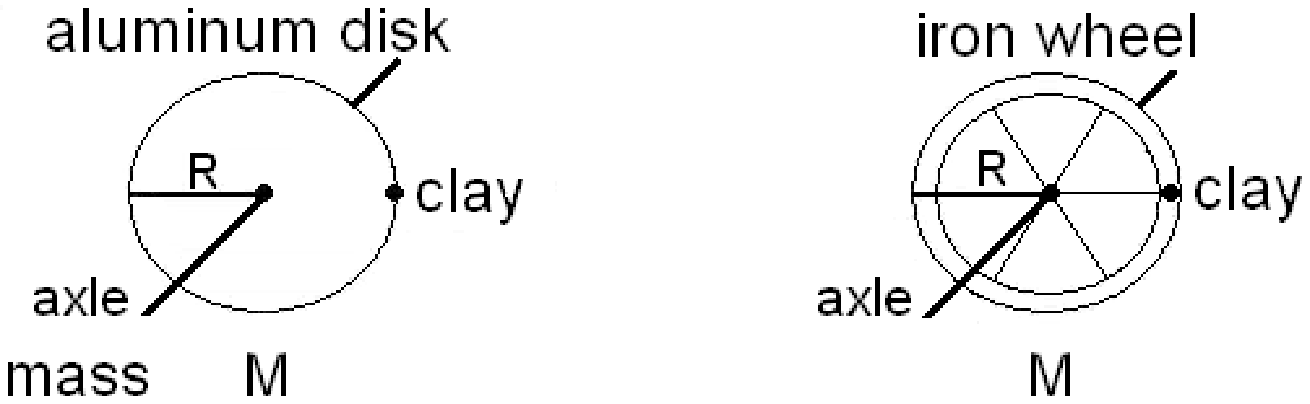}
\end{center}
\noindent
{\bf  Explain:}
\vspace{0.1in}

\item
Consider the net torque acting on the disk+clay and wheel+clay systems, about a point on its own axle.
Which one of the following statements is \underline {true}? 
\begin{itemize}
\item[(a)]
The net torque is greater for the disk+clay system.
\item[(b)]
The net torque is greater for the wheel+clay system.
\item[(c)]
Which system has a greater net torque depends on the actual numerical values of $R$ and $M$.
\item[(d)]
There is no net torque on either system.
\item[(e)]
The net torques on both systems are equal and non-zero.
\end{itemize}
\noindent
{\bf  Explain:}
\vspace{0.1in}

\item
Which one of the following statements about their angular accelerations is \underline {true}?
\begin{itemize}
\item[(a)]
The angular acceleration is greater for the disk+clay system.
\item[(b)]
The angular acceleration is greater for the wheel+clay system.
\item[(c)]
Which system has a greater angular acceleration depends on the actual numerical values of $R$ and $M$.
\item[(d)]
There is no angular acceleration for either system.
\item[(e)]
The angular accelerations of both systems are equal and non-zero.
\end{itemize}
\noindent
{\bf  Explain:}
\vspace{0.1in}

\item
Which one of the following statements about their maximum angular velocities is \underline {true}?
\begin{itemize}
\item[(a)]
The maximum angular velocity is greater for the disk+clay system.
\item[(b)]
The maximum angular velocity is greater for the wheel+clay system.
\item[(c)]
Which object has a greater maximum angular velocity is determined by the actual numerical values of $R$ and $M$.
\item[(d)]
The maximum angular velocities of both systems are equal and non-zero.
\item[(e)]
There is no angular velocity for either system so the question of a maximum value does not arise.
\end{itemize}
\noindent
{\bf  Explain:}
\vspace{0.1in}

\pagebreak

\item
An aluminum disk and a plastic disk have different radii but the same mass as shown below. 
Each disk is free to rotate about its own fixed horizontal frictionless axle.
Both disks are initially at rest.
\underline{Identical} small lumps of clay are attached to their rims as shown in the figure.
Consider the net torque acting on each disk+clay system, about a point on its own axle.
Which one of the following statements is \underline {true}? 
\begin{center}
\includegraphics[height=1.9in]{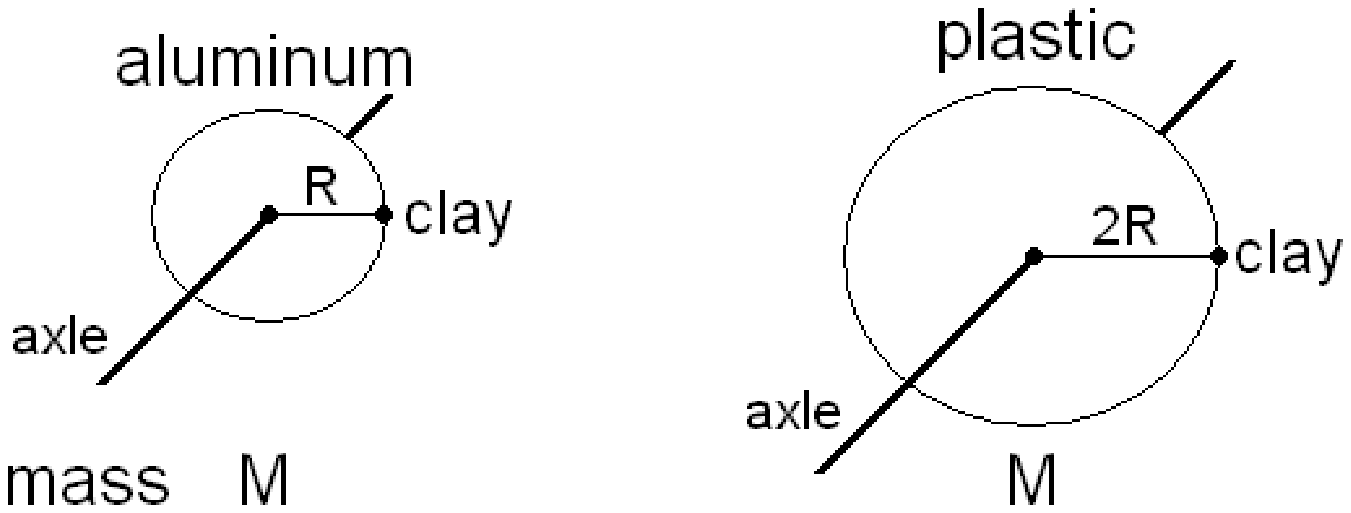}
\end{center}
\begin{itemize}
\item[(a)]
The net torque is greater for the system with the plastic disk.
\item[(b)]
The net torque is greater for the system with the aluminum disk.
\item[(c)]
Which system has a greater net torque depends on the actual numerical values of the radii of the disks.
\item[(d)]
There is no net torque on either system.
\item[(e)]
The net torques on both systems are equal and non-zero.
\end{itemize}
\noindent
{\bf  Explain:}
\vspace{0.1in}

\item
A copper disk and a plastic disk have different radii and masses as shown below. 
Each disk is free to rotate about its own fixed horizontal frictionless axle.
Both disks are initially at rest.
\underline{Identical} small lumps of clay are attached to their rims as shown in the figure.
Consider the net torque acting on each disk+clay system, about a point on its own axle.
Which one of the following statements is \underline {true}? 
\begin{center}
\includegraphics[height=1.9in]{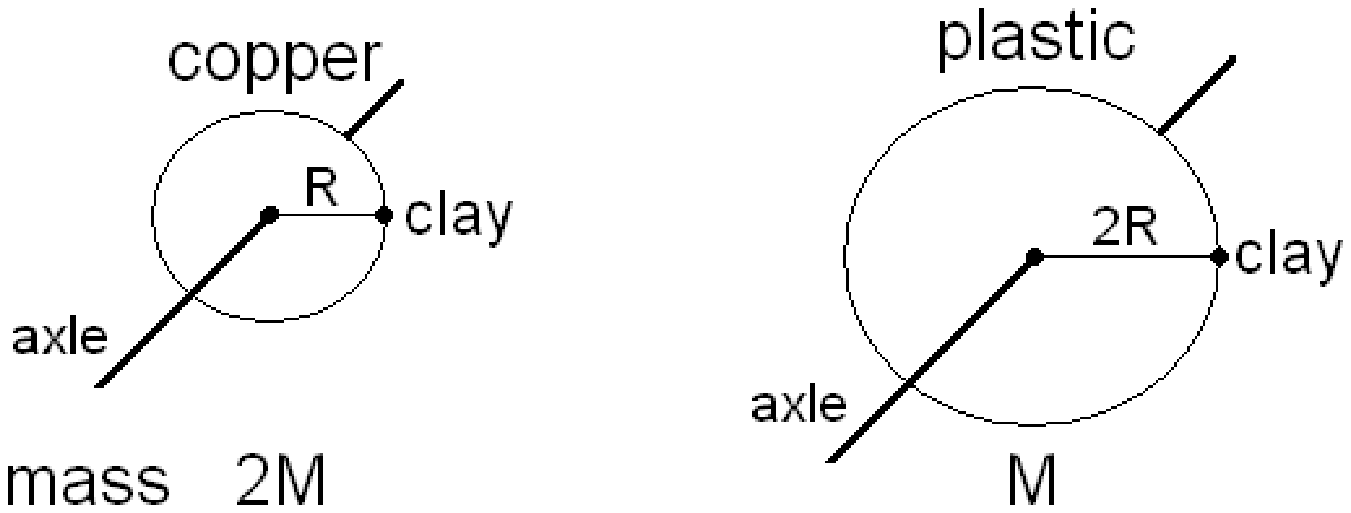}
\end{center}
\begin{itemize}
\item[(a)]
The net torque is greater for the system with the plastic disk.
\item[(b)]
The net torque is greater for the system with the copper disk.
\item[(c)]
Which system has a greater net torque depends on the actual numerical values of the masses and the radii of the disks.
\item[(d)]
There is no net torque on either system.
\item[(e)]
The net torques on both systems are equal and non-zero.
\end{itemize}

\noindent
{\bf  Explain:}
\vspace{0.1in}

\pagebreak

{\bf {\Large $\bullet$} \underline{Setup for the next two questions}}\\
{\bf 
Two copper disks of different thicknesses have the same radius but different masses as shown below. 
Each disk is free to rotate about its own fixed horizontal frictionless axle.
Both disks are initially at rest.
\underline{Identical} small lumps of clay are attached to their rims as shown in the figure.
}
\begin{center}
\includegraphics[height=.99in]{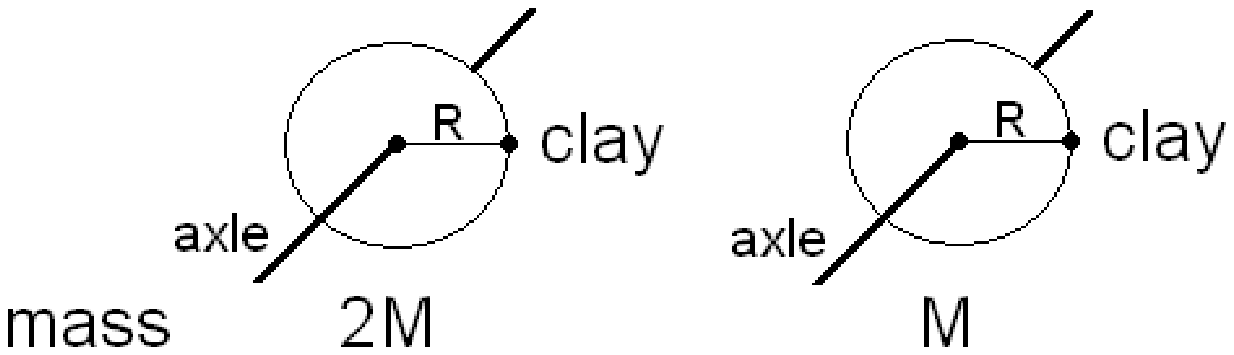}
\end{center}

\item
Consider the net torque acting on each disk+clay system, about a point on its own axle.
Which one of the following statements is \underline {true}? 
\begin{itemize}
\item[(a)]
The net torque is greater for the system in which the disk has larger mass.
\item[(b)]
The net torque is greater for the system in which the disk has smaller mass.
\item[(c)]
Which system has a greater net torque depends on the actual numerical values of their masses.
\item[(d)]
There is no net torque on either system.
\item[(e)]
The net torques on both systems are equal and non-zero.
\end{itemize}

\noindent
{\bf  Explain:}
\vspace{0.1in}

\item
Which one of the following statements about their angular accelerations is \underline {true}? 
\begin{itemize}
\item[(a)]
The angular acceleration is greater for the system in which the disk has larger mass.
\item[(b)]
The angular acceleration is greater for the system in which the disk has smaller mass.
\item[(c)]
Which system has a greater angular acceleration depends on the actual numerical values of their masses.
\item[(d)]
There is no angular acceleration for either system.
\item[(e)]
The angular accelerations of both systems are equal and non-zero.
\end{itemize}

\noindent
{\bf  Explain:}
\vspace{0.1in}

\item
A wheel of mass $M$ and radius $R$ is free to rotate about a fixed horizontal axle.
A small lump of clay of mass $m$ is attached to its rim as shown in the figure.
Consider the magnitude of the net torque $\tau$ acting on the 
wheel+clay system about a point on the axle shown in the figure below, and let $g$ be the magnitude of the acceleration due to gravity.
Which one of the following statements is \underline {true}? 
\begin{center}
\includegraphics[height=.99in]{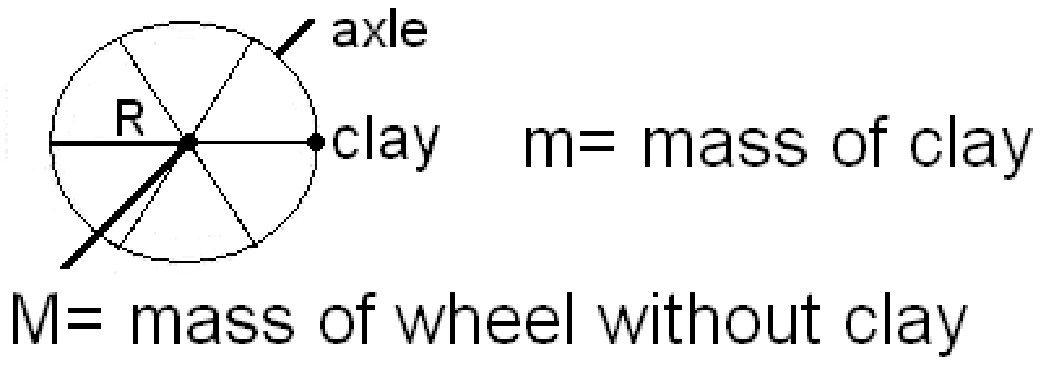}
\end{center}
\begin{itemize}
\item[(a)]
$\tau=mgR$. 
\item[(b)]
$\tau=(m+M)gR$.
\item[(c)]
$\tau=(m+M/2)gR$.
\item[(d)]
$\tau=(m+M)g/R$.
\item[(e)]
None of the above.
\end{itemize}

\noindent
{\bf Explain:}

\end{enumerate}

\end{document}